\documentclass[twocolumn]{aastex63}
\usepackage{color,amsmath,multirow}

\newcommand{\vrad}{{$v_{\mathrm{r}}$}~}
\newcommand{\kms}{{km s$^{-1}$}~}
\newcommand{\kmsend}{{km s$^{-1}$}}

\accepted{September 29, 2020}
\submitjournal{ApJ}

\shorttitle{Kinematics of M85 GCs}
\shortauthors{Ko et al.}

\begin{document}

\title{Mysterious Globular Cluster System of the Peculiar Massive Galaxy M85}

\correspondingauthor{Myung Gyoon Lee}
\email{mglee@astro.snu.ac.kr}

\author{Youkyung Ko}
\affiliation{Korea Astronomy and Space Science Institute, 776 Daedeok-daero, Yuseong-Gu, Daejeon 34055, Korea}
\affil{Astronomy Program, Department of Physics and Astronomy, Seoul National University, 1 Gwanak-ro, Gwanak-gu, Seoul 08826, Korea}

\author{Myung Gyoon Lee}
\affiliation{Astronomy Program, Department of Physics and Astronomy, Seoul National University, 1 Gwanak-ro, Gwanak-gu, Seoul 08826, Korea}

\author{Hong Soo Park}
\affiliation{Korea Astronomy and Space Science Institute, 776 Daedeok-daero, Yuseong-Gu, Daejeon 34055, Korea}

\author{Jubee Sohn}
\affiliation{Smithsonian Astrophysical Observatory, 60 Garden Street, Cambridge, 02138, USA}

\author{Sungsoon Lim}
\affiliation{University of Tampa, 401 West Kennedy Boulevard, Tampa, FL 33606, USA}

\author{Narae Hwang}
\affiliation{Korea Astronomy and Space Science Institute, 776 Daedeok-daero, Yuseong-Gu, Daejeon 34055, Korea}

\author{Byeong-Gon Park}
\affiliation{Korea Astronomy and Space Science Institute, 776 Daedeok-daero, Yuseong-Gu, Daejeon 34055, Korea}




\begin{abstract}
We present a study on stellar population and kinematics of globular clusters (GCs) in the peculiar galaxy M85. We obtain optical spectra of 89 GCs at 8 kpc $< R <$ 160 kpc using the MMT/Hectospec. We divide them into three groups, blue/green/red GCs (B/G/RGCs), with their $(g-i)_0$ colors. All GC subpopulations have mean ages of 10 Gyr, but showing differences in metallicities. The BGCs and RGCs are the most metal-poor ([Z/H] $\sim -1.49$) and metal-rich ([Z/H] $\sim -0.45$), respectively, and the GGCs are in between. We find that the inner GC system exhibits a strong overall rotation that is entirely due to a disk-like rotation of the RGC system. The BGC system shows little rotation. The GGCs show kinematic properties clearly distinct among the GC subpopulations, having higher mean velocities than the BGCs and RGCs and being aligned along the major axis of M85. This implies that the GGCs have an origin different from the other GC subpopulations. The rotation-corrected velocity dispersion of the RGC system is much lower than that of the BGC system, indicating the truncation of the red halo of M85. The BGCs have a flat velocity dispersion profile out to $R$ = 67 kpc, reflecting the dark matter extent of M85. Using the velocity dispersion of the BGC system, we estimate the dynamical mass of M85 to be $3.8 \times 10^{12} M_{\odot}$. We infer that M85 has undergone merging events lately, resulting in the peculiar kinematics of the GC system.
\end{abstract}





\section{Introduction}

Massive elliptical galaxies have formed via continuous mergers in the hierarchical galaxy formation model \citep{tt72}. In this scenario, merger remnant galaxies show a snapshot of evolutionary stages between disk and elliptical galaxies. Many studies reported that nearby ellipticals have fine structures produced during past mergers \citep[e.g.][]{sch82, duc15}.

M85 (NGC 4382, VCC 798) is one of the nearby merger remnant galaxies that show interesting merging features.
Because of these merger remnant features, the morphological type of M85 has been uncertain.
\citet{bst85} and \citet{dev91} classified M85 as an S0pec because of its disk-like structures in addition to its prominent bulge.
However, \citet{kor09} suggested that its morphological type is an E2, not an S0, because the $a_4$ profile derived from ellipse fitting does not indicate any disky structure at the radial range of $26 \arcsec < R < 221\arcsec$, corresponding to 2 kpc $ < R < $ 19 kpc at a distance to M85 of 17.9 Mpc \citep{bla09}, where the light excess in the surface brightness profile appears.

Elliptical galaxies are divided into two groups in general, called as `E-E dichotomy' \citep[][and references therein]{kor09}: (1) giant ellipticals ($M_V < -21.5$ mag) which generally have cuspy cores and boxy-distorted isophotes, rarely rotate, and have mostly old stars, and (2) normal and dwarf ellipticals ($M_V > -21.5$ mag) that lack cores, but have extra light at the center, strongly rotate with disky-distorted isophotes, and have younger stars.
Interestingly, M85 is an exceptional case in this dichotomy.
M85 is classified as a giant elliptical galaxy according to its brightness, $M_V^T = -22.52$ mag \citep{kor09}. It has a core and boxy isophotes within 1$\arcsec$ from the galaxy center \citep{fer06}.
These are all general properties of giant ellipticals.
However, M85 also shows unusual properties that giant ellipticals rarely have.
Several studies found that the nucleus of M85 is as young as a few Gyrs based on the spectroscopic analysis \citep{ffi96, tf02, ko18}.
In addition, \citet{ems07} classified M85 as a fast rotator with a projected specific angular momentum of $\lambda_{R} = 0.155$ at one effective radius ($R_e = 67\arcsec$).
These properties are related to post merger events.

All these studies about the stellar light of M85 focused on the central region within one effective radius.
There have been several studies that investigated a wider region of M85 using globular clusters (GCs) that are a useful tool to study galaxy halo structures \citep{pen06, chi11, tra14}.
These studies found that the GCs in M85 also show peculiar properties like the central stars in M85.
In general, GCs in massive early-type galaxies show a bimodal optical color distribution, which indicates the existence of two GC subpopulations: old metal-poor (blue) and old metal-rich (red) GCs.
However, the GCs at $R < 2\arcmin$ (10 kpc) in M85 do not clearly show a bimodal color distribution \citep{pen06}.
This implies the presence of intermediate-age GCs, indicating that their host galaxies have experienced mergers accompanying intense star formation a few Gyrs ago.
\citet{chi11} and \citet{tra14} confirmed the existence of the intermediate-age GC populations in M85 using the combination of optical and $K$-band photometry.
These previous studies covered only the central region at $R < 2\arcmin$.

In this context, we performed a wide-field photometric survey of the GCs in M85, covering a $1\arcdeg \times 1\arcdeg$ field \citep[][hereafter Paper I]{ko19}.
We identified 1318 GC candidates in the survey region, and found that the radial extent of the GC system of M85 is as large as $R = 20\arcmin$ (104 kpc).
Also we detected a number of intermediate-color GC candidates in the central region ($R < 2\arcmin$), which is consistent with the previous study \citep{pen06}.
As a follow-up, \citet[][hereafter Paper II]{ko18} measured the ages and metallicities of 20 GCs in M85 using the optical spectra obtained with the Gemini/GMOS. We found that 55\% of the GCs have mean ages of about 4 Gyr, much younger than typical GCs.
In addition, we detected a strong disk-like rotation of the GC system with a rotation amplitude of 148 \kmsend.
However, these results are needed to be supplemented with a larger sample because this spectroscopic survey covers only the small central region at $R < 3\arcmin$ (16 kpc) although the M85 GC system is extended to $R = 20\arcmin$ according to the photometric survey (Paper I).

In this study, we present a wide-field spectroscopic survey of the GCs in M85 to investigate the physical properties of the GCs in the outskirts of M85. We cover $R < 30\arcmin$ (156 kpc) using the Hectospec on the 6.5 m MMT.
To date this GC survey covers the widest area around M85.
This paper is organized as follows. 
We briefly describe the spectroscopic target selection, observation, and data reduction in Section 2.
In Section 3, we identify genuine GCs, and investigate the stellar population and kinematic properties of GC subpopulations of M85.
We discuss the peculiarity of the M85 GC system, and investigate the dark matter extent of M85 as well as dynamical mass estimation in Section 4.
We summarize the results in Section 5.

\section{Observation and Data Reduction}

	\subsection{Target selection and Spectroscopic Observation}

\begin{deluxetable*}{c c c c c c c}
\tablecaption{Observation Log for the MMT/Hectospec Run \label{tab:obs.field}}
\tablewidth{0pt}
\tablehead{
\colhead{Mask Name} & \colhead{$\alpha$ (J2000)} & \colhead{$\delta$ (J2000)} & \colhead{N$^{a}$}
& \colhead{Exp. time} & \colhead{Seeing} & \colhead{Date(UT)}
}
\startdata
M85-B1 & 12:25:24.74 & +18:10:21.3 & 256 & 5 $\times$ 1440 s & 1\farcs3 & Mar 7, 2016 \\
M85-B2 & 12:25:20.08 & +18:05:08.9 & 260 & 5 $\times$ 1440 s & 0\farcs9 & Mar 16, 2016 \\
M85-F1 & 12:25:26.52 & +18:07:07.6 & 250 & 5 $\times$ 1800 s & 1\farcs2 & Mar 17, 2016 \\
\enddata
\tablecomments{$^{a}$ Number of object fibers among 300 fibers in each field. The remaining fibers are assigned to sky regions.}
\end{deluxetable*}

	\begin{figure}[t]
\epsscale{1}
\includegraphics[width=\columnwidth]{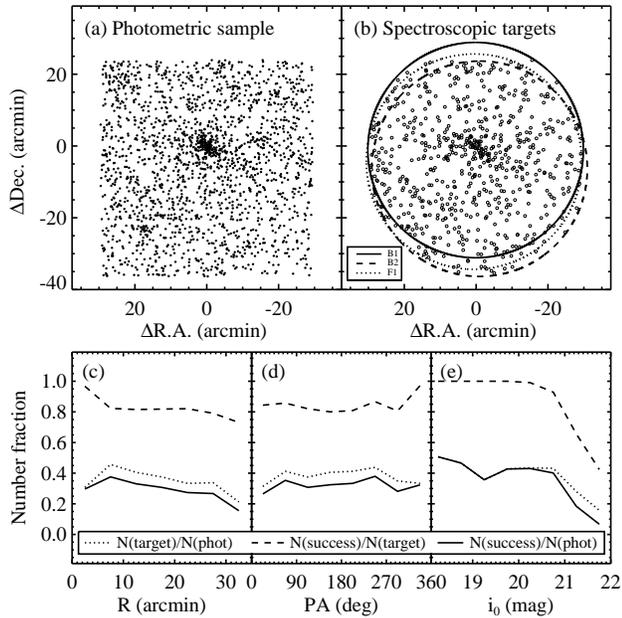}
\caption{(a) Spatial distribution of GC candidates with $18 < i_0 < 22$ identified by Paper I. (0, 0) indicates the center of M85.
(b) Same as (a), but for spectroscopic targets. Each large circle represents a field of view of MMT/Hectospec with a 1$^{\circ}$ diameter.
The solid-line, dashed-line, and dotted-line circles represent the M85-B1, M85-B2, and M85-F1 field configurations.
(c) Number fractions of the targets with velocity measurements to the photometric sample (solid line) and to the spectroscopic targets (dashed line) as a function of galactocentric distance from M85.
(d) Same as (c), but for as a function of position angle.
\label{fig:target_spatial}}
	\end{figure}
	
We selected the spectroscopic targets from the photometric sample of GC candidates around M85 in Paper I.
The GC candidates were identified as point-like sources in the $ugi$ band images taken with the MegaCam at the 3.8 m Canada-France-Hawaii Telescope.
We used the $(u-g)_0$ and $(g-i)_0$ color combination to select the GC candidates. 
The magnitude range for the target selection was set to be $18 < i_0 < 22$ in order to minimize the contamination of foreground stars.
The spatial distribution of these GC candidates is shown in Figure \ref{fig:target_spatial}(a).
We assigned fibers to a total of 645 GC candidates for spectroscopic observations ({Figure \ref{fig:target_spatial}(b)}).
In addition to the GC candidates, we obtained spectra of the M85 nucleus and a hypercompact star cluster M85-HCC1 discovered by \citet{san15}.

We carried out spectroscopic observation using the Hectospec \citep{fab05} mounted on the 6.5 m MMT (program ID: 2016A-UAO-G4, PI: Youkyung Ko) during March 2016.
We selected a 270 mm$^{-1}$ grating with a dispersion of 1.2 \AA~ pixel$^{-1}$, covering the wavelength range of 3650 -- 9200 \AA.
Three different configurations with a slight offset, covering the $R < 30\arcmin$ region around M85, were made, as shown in {Figure \ref{fig:target_spatial}(b)}.
We used the exposure time of 7200 s (five times of 1440 s) for each of the two configurations (M85-B1 and B2) to cover bright targets, and we used the longer exposure time of 9000 s (five times of 1800 s) for one configuration (M85-F1) to cover fainter targets.
The seeing ranged from 0\farcs9 to 1\farcs3 during the observations.
The field coordinates and exposure times are given in {Table \ref{tab:obs.field}}.

{We calculated the completeness of the mask design, defined as the ratio between the targets on which fibers are allocated and photometric samples ($N({\rm target})/N({\rm phot})$), as functions of galactocentric distance, position angle, and $i$-band magnitude ({Figure \ref{fig:target_spatial}(c)-(e)}). We found that the completeness is almost constant ($\sim$ 30\%) over entire radial and azimuthal ranges, indicating that there is no bias on the target allocation along the location. The completeness is constant for the bright sources with $i_0 < 21$ mag, but decreases for the fainter sources.}

{In addition, we compared the color distribution of the photometric samples with that of the targets on which fibers are assigned. The fiber allocation rate is constant ($\sim$ 30\%) in the color range of $0.55 < (g-i)_0 < 1.2$ that corresponds to the GC color, indicating that there is no bias on the target selection.}

	\subsection{Data Reduction and Radial Velocity Measurements}

We used version 2.0 of the HSRED reduction pipeline\footnote{This is an updated reduction pipeline originally developed by Richard Cool; more details can be found at http://www.mmto.org/node/536.} for data reduction.
It includes bias and dark correction, flat-fielding, aperture extraction of spectra, and wavelength calibration.
Flux calibration was done following the methods described by \citet{fab08}.
Most of the faint targets with $i > 21.0$ mag have low signal-to-noise ratios ($S/N <$ 5). 
The median signal-to-noise ratio of the spectra of GC candidates with $i < 21.0$ mag at 5000 \AA~ is $S/N \sim 10$.

We estimated heliocentric radial velocities of spectroscopic targets using the xcsao task in the IRAF RVSAO package \citep{km98}.
The prominent absorption lines in the wavelength range of 3800 -- 5400 \AA~ were used to apply the cross-correlation method \citep{td79}.
The RVSAO package presents several radial velocity templates such as spectra of an A star, M31 GCs, elliptical and spiral galaxies.
We used 10 templates, and matched the targets with \vrad $>$ 3000 \kms and \vrad $<$ 3000 \kms with galaxy and GC templates, respectively.
For 115 of the 645 targets, we could not derive reliable radial velocities because of low signal-to-noise ratios ($S/N <$ 5).
In addition, we excluded 21 targets fainter than the luminosity of the galaxy light within the fiber, using the surface brightness profile of M85 from \citet{kor09}.

We also measured radial velocities of the M85 nucleus and the M85-HCC1, and compared the measurements with the results in previous studies.
The radial velocity of the M85 nucleus is derived to be \vrad = 695 $\pm$ 16 \kmsend, which is smaller than those in several previous studies, \vrad = 729 -- 760 \kms \citep[][Paper II]{smi00, gav04}.
In the case of M85-HCC1, the radial velocity is measured to be \vrad = 655 $\pm$ 7 \kmsend, which is consistent with the result from the SDSS DR15 \citep{agu19}, \vrad = 664 $\pm$ 5, within uncertainties.
{We did not add any offset value to the radial velocities we measured because the velocity measurements for the point source M85-HCC1 agree well.}

We calculated the spectroscopic success rate defined as the number ratio between the targets of which radial velocities are well derived and the parent photometric sample ($N({\rm success})/N({\rm phot})$) as a function of galactocentric distance from M85 ({Figure \ref{fig:target_spatial}(c)}).
We found that the success rates are almost constant as $\sim$ 30\% for the whole radial ranges.
For comparison, we also calculated the number fraction of the targets with velocity measurements to the spectroscopic targets on which fibers were allocated ($N({\rm success})/N({\rm target})$). This fraction does not vary significantly with distance from M85.
The azimuthal variations of $N({\rm success})/N({\rm phot})$ and $N({\rm success})/N({\rm target})$ are similar to their radial variations ({Figure \ref{fig:target_spatial}(d)}).
Therefore, we conclude that there is no bias in the velocity measurements along the target location.
{In addition, we checked the $N({\rm success})/N({\rm phot})$ and $N({\rm success})/N({\rm target})$ as a function of $i$-band magnitude ({Figure \ref{fig:target_spatial}(e)}). We found that the radial velocities of all the bright targets with $i_0 < 21$ mag are successfully measured with the mean spectroscopic success rate of 40\%.
}

	\begin{figure}[t]
\epsscale{1}
\includegraphics[width=\columnwidth]{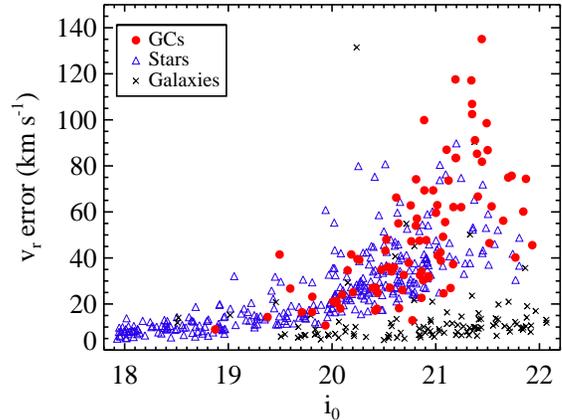}
\caption{Radial velocity errors vs. dereddened $i$-band magnitudes for the spectroscopic targets. The red filled circles, blue open triangles, and black crosses represent the GCs, foreground stars, and background galaxies confirmed in this study.
\label{fig:evr_imag}}	
	\end{figure}

{Figure \ref{fig:evr_imag}} shows radial velocity uncertainties versus dereddened $i$-band magnitudes for the spectroscopic targets classified into GCs, foreground stars, and background galaxies (see Section 3.1).
On average, brighter sources have smaller velocity uncertainties than fainter sources.
The mean radial velocity uncertainty of the sources with $i_0 < 19.5$ mag, which are mostly foreground stars, is 11 \kmsend.
Most of the GCs have $i$-band magnitudes of $19.5 < i_0 < 21$ mag, where the mean velocity uncertainty ranges from 15 to 30 \kmsend.
The velocity uncertainties of background galaxies are mostly smaller than 20 \kmsend, regardless of their luminosity.
This is because the velocity measurements for the faint galaxies were based on emission lines that are much stronger and narrower than absorption lines.

	\begin{figure}[t]
\epsscale{1}
\includegraphics[width=\columnwidth]{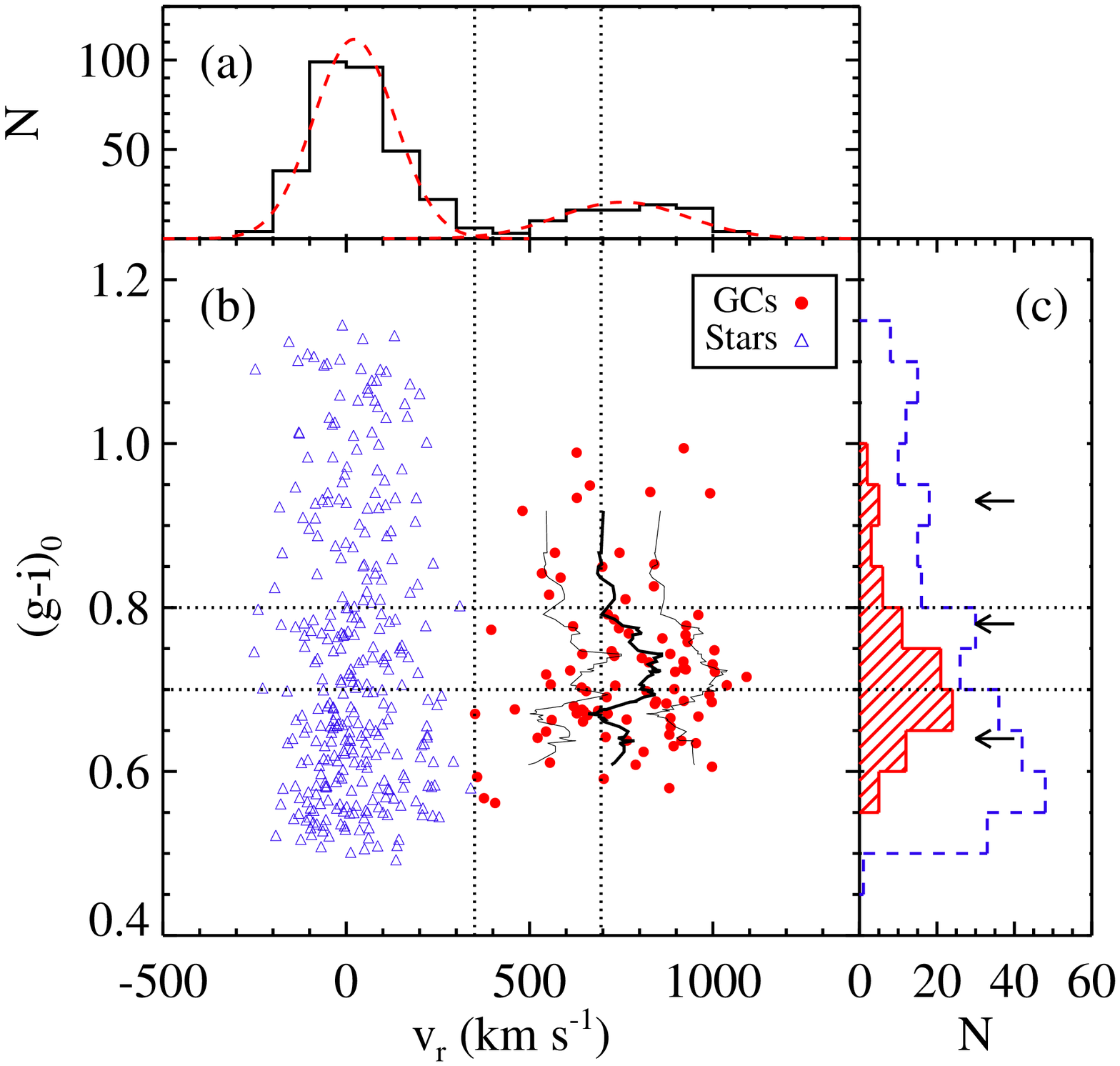}
\caption{(a) Radial velocity distribution of the spectroscopic targets with \vrad $<$ 3000 \kmsend. The dashed line indicates the GMM result for their radial velocity distribution.
(b) $(g-i)_0$ colors vs. radial velocities of the foreground stars (blue open triangles) and GCs (red filled circles) confirmed in this study. {The thick solid line indicates rolling averages of velocities as a function of $(g-i)_0$ color with moving bins of $N = 12$. The thin solid lines at either side show the standard deviation from the mean values.}
(c) $(g-i)_0$ color distribution of the foreground stars (dashed line) and GCs (solid line). {The horizontal arrows indicate three peaks in the $(g-i)_0$ color distribution of the parent photometric sample derived from the GMM test (Paper I).}
The two dotted vertical lines indicate the radial velocity criterion for dividing the targets into GCs and foreground stars (\vrad = 350 \kmsend) and the radial velocity of the M85 nucleus derived in this study (\vrad = 696 \kmsend).
The two horizontal lines are the color criteria for dividing the GCs into blue, green, and red ($(g-i)_0$ = 0.7 and 0.8).
\label{fig:v-color}}	
	\end{figure}

\section{Results}

	\subsection{GC Selection and Subpopulations}

\begin{deluxetable*}{l l l c c c c c}
\setlength{\tabcolsep}{12pt}
\tabletypesize{\footnotesize}
\tablecaption{Spectroscopic and photometric properties of the GCs confirmed in this study \label{tab:gc.prop}}
\tablehead{
\colhead{ID} & \colhead{$\alpha$ (J2000)} & \colhead{$\delta$ (J2000)} & \colhead{$i^a$} & \colhead{$(g-i)^a$} & \colhead{$C^b$} & \colhead{\vrad} & \colhead{Class$^c$} \\
\colhead{} & \colhead{(deg)} & \colhead{(deg)} & \colhead{(mag)} & \colhead{(mag)} & \colhead{(mag)} & \colhead{(\kmsend)} & \colhead{}
}
\startdata
GC01 & 185.892609 & 18.149979 & 19.429 $\pm$  0.002 &  0.822 $\pm$  0.003 &  0.02 & 395 $\pm$ 14 & GGC \\
GC02 & 185.961578 & 18.098152 & 21.175 $\pm$  0.005 &  0.792 $\pm$  0.009 &  0.12 & 884 $\pm$ 74 & GGC \\
GC03 & 185.973221 & 17.820648 & 20.941 $\pm$  0.005 &  0.840 $\pm$  0.008 &  0.14 & 960 $\pm$ 69 & GGC \\
GC04 & 186.040955 & 18.078588 & 20.691 $\pm$  0.004 &  0.754 $\pm$  0.006 &  0.08 & 734 $\pm$ 55 & GGC \\
GC05 & 186.095642 & 18.139322 & 20.314 $\pm$  0.003 &  0.734 $\pm$  0.005 &  0.07 & 997 $\pm$ 39 & BGC \\
\enddata
\tablecomments{{Table \ref{tab:gc.prop}} is published in its entirety in the electronic edition. The five sample GCs are shown here as guidance for the table’s form and content.}
\tablenotetext{a}{CFHT/MegaCam AB magnitudes.}
\tablenotetext{b}{The inverse concentration index $C$ defined as the $i$-band magnitude differences between 4- and 8-pixel-diameter aperture photometry.}
\tablenotetext{c}{Classifications are BGC (blue GC with $(g-i)_0 < 0.7$), GGC (green GC with $0.7 < (g-i)_0 < 0.8$), and RGC (red GC with $(g-i)_0 > 0.8$.}
\end{deluxetable*}

\begin{deluxetable*}{l l l c c c c}
\setlength{\tabcolsep}{12pt}
\tabletypesize{\footnotesize}
\tablecaption{Spectroscopic and photometric properties of foreground stars and background galaxies in the Hectospec field of M85\label{tab:star_gal.prop}}
\tablewidth{0pt}
\tablehead{
\colhead{ID} & \colhead{$\alpha$ (J2000)} & \colhead{$\delta$ (J2000)} & \colhead{$i^a$} & \colhead{$(g-i)^a$} & \colhead{$C^b$} & \colhead{\vrad}\\
\colhead{} & \colhead{(deg)} & \colhead{(deg)} & \colhead{(mag)} & \colhead{(mag)} & \colhead{(mag)} & \colhead{(\kmsend)}
}
\startdata
Star001 & 185.835114 & 18.226645 & 20.423 $\pm$  0.004 &  0.689 $\pm$  0.006 & --0.02 & --146 $\pm$ 40 \\
Star002 & 185.835785 & 18.116449 & 20.406 $\pm$  0.004 &  0.849 $\pm$  0.006 & --0.02 & 0 $\pm$ 22 \\
Star003 & 185.839340 & 18.155159 & 20.080 $\pm$  0.002 &  1.005 $\pm$  0.004 &  0.00 & 105 $\pm$ 20 \\
Star004 & 185.846497 & 18.014795 & 20.407 $\pm$  0.003 &  0.629 $\pm$  0.005 & 0.00 & 340 $\pm$ 37 \\
Star005 & 185.849579 & 18.077505 & 20.706 $\pm$  0.004 &  0.789 $\pm$  0.006 &  0.02 & 169 $\pm$ 64 \\
\hline
Galx001 & 185.882706 & 18.143103 & 21.300 $\pm$  0.006 &  0.554 $\pm$  0.009 &  0.32 & 46086 $\pm$ 6 \\
Galx002 & 185.896484 & 17.997314 & 21.896 $\pm$  0.011 &  0.618 $\pm$  0.015 &  0.25 & 31115 $\pm$ 10 \\
Galx003 & 185.900116 & 17.980145 & 21.834 $\pm$  0.010 &  0.489 $\pm$  0.013 &  0.31 & 46201 $\pm$ 11 \\
Galx004 & 185.904617 & 18.043806 & 19.892 $\pm$  0.002 &  0.955 $\pm$  0.004 &  0.38 & 34970 $\pm$ 9 \\
Galx005 & 185.906403 & 18.110588 & 21.971 $\pm$  0.011 &  0.832 $\pm$  0.017 &  0.22 & 46142 $\pm$ 11 \\
\enddata
\tablecomments{{Table \ref{tab:star_gal.prop}} is published in its entirety in the electronic edition. The five sample stars and galaxies are shown here as guidance for the table’s form and content.}
\tablenotetext{a}{CFHT/MegaCam AB magnitudes.}
\tablenotetext{b}{The inverse concentration index $C$ defined as the difference between 4- and 8-pixel-diameter $i$-band aperture-corrected magnitudes.}
\end{deluxetable*}

We classified the observed targets into GCs, foreground stars, and background galaxies.
First, there are 110 background galaxies with \vrad $>$ 3000 \kmsend.
The radial velocity distribution of the galaxies in the Virgo Cluster field shows a clear separation at \vrad = 3000 \kms \citep{kim14}.
We adopted this criterion to divide the targets into the objects bound to the Virgo Cluster and background galaxies.
The radial velocities of the background galaxies in this sample range from 16883 \kms to 151120 \kmsend, which are much higher than the mean velocity of the Virgo Cluster galaxies in the survey region, \vrad $\sim$ 1056 \kms \citep{kim14}.

The targets with \vrad $<$ 3000 \kms are either M85 GCs or foreground stars.
The radial velocity distribution of these targets shows two peaks clearly at \vrad $\sim$ 0 \kms and 700 \kms ({Figure \ref{fig:v-color}(a)}) corresponding to foreground stars and M85 GCs, respectively.
To decompose these two populations, we performed the Gaussian Mixture Modeling \citep[GMM;][]{mg10}, assuming bimodal distribution with different variances.
The $p$ and $D$ values derived from the GMM indicate the probability of the unimodal distribution and the peak separation relative to the Gaussian width, respectively.
In this case, the $p$ value is smaller than 0.0001 and the $D$ value is 5.01, which means that the input radial velocity distribution is not unimodal and has a clear peak separation.

The mean radial velocities of the two populations are \vrad = 22 $\pm$ 6 \kms and 754 $\pm$ 23 \kmsend, and the Gaussian widths are 111 $\pm$ 4 \kms and 174 $\pm$ 15 \kmsend, respectively.
According to the GMM results, we adopted the point where two Gaussians cross (\vrad = 350 \kmsend) as the radial velocity criterion for dividing the targets into GCs and foreground stars.
We consider 89 targets with \vrad $>$ 350 \kms and 310 targets with \vrad $<$ 350 \kms to be GCs and foreground stars, respectively.
{Table \ref{tab:gc.prop}} and {Table \ref{tab:star_gal.prop}} list photometric properties and radial velocities of GCs and the contaminants (foreground stars and background galaxies), respectively.

{Figure \ref{fig:v-color}(b)} shows the $(g-i)_0$ colors and radial velocities of GCs and foreground stars confirmed in this study.
The magnitudes are based on the CFHT/MegaCam AB system.
We used a foreground extinction value of $E(B-V) = 0.024$ mag ($E(g-i) = 0.049$ mag) for M85 \citep{sf11}.
The foreground stars show a clear sequence at \vrad = 0 \kms with a broad color range of $(g-i)_0 = 0.5 - 1.15$, while the GCs have a narrower color range of $(g-i)_0 = 0.55 - 1.0$ at \vrad = 750 \kmsend.
{Figure \ref{fig:v-color}(c)} displays that the $(g-i)_0$ color distribution of the GCs shows a dominant peak at $(g-i)_0 = 0.675$ and a much weaker peak at $(g-i)_0 = 0.925$, which is consistent with those of the parent sample (Paper I).
The foreground stars have a strong blue peak at $(g-i)_0 = 0.575$ and a red tail in the $(g-i)_0$ color distribution.


We calculated rolling averages of velocities as a function of $(g-i)_0$ color with moving bins of $N = 12$, and note that intermediate-color GCs with $0.7 \lesssim (g-i)_0 \lesssim 0.8$ have radial velocities higher than the other GCs ({Figure \ref{fig:v-color}(b)}).
It has been known that the GCs in M85 show a trimodal color distribution, while the GCs in massive early-type galaxies often show a bimodal color distribution \citep{pen06}. Paper I also suggested a possibility that M85 has an intermediate-color GC population with a mean color of $(g-i)_0 \sim 0.78$ based on the GMM tests despite a high uncertainty of its number fraction.
Because we identified a velocity peculiarity of these green GCs (GGCs) with $0.7 < (g-i)_0 < 0.8$, we consider them as a separate population in addition to blue GCs (BGCs) with $(g-i)_0 < 0.7$ and red GCs (RGCs) with $(g-i)_0 > 0.8$.
The numbers of the BGCs, GGCs, and RGCs in M85 are 41, 32, and 16. The subpopulation information of the GCs is also listed in {Table \ref{tab:gc.prop}}.

	\begin{figure}[t]
\epsscale{1}
\includegraphics[width=\columnwidth]{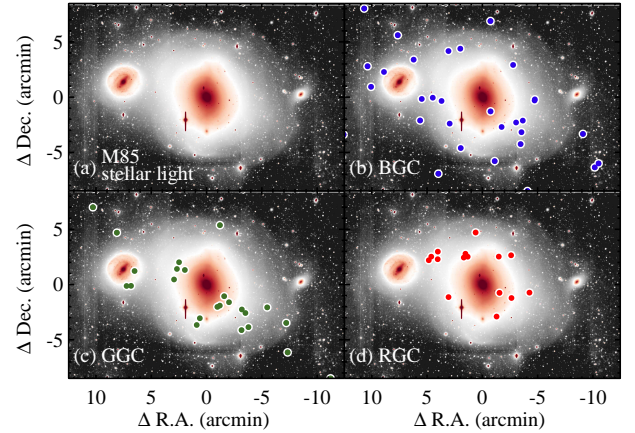}
\caption{(a) $g$-band image of a $25\arcmin \times 15\arcmin$ field around M85 taken with CFHT/MegaCam (Paper I), including NGC 4394 (east) and IC 3292 (west). (b) Spatial distribution of BGCs overlaid on the $g$-band image. (c)-(d) Same as (b), but for GGCs and RGCs. \label{fig:image_bgr}}
	\end{figure}
	
{Figure \ref{fig:image_bgr}} shows spatial distributions of the GC subpopulations within a $25\arcmin \times 15\arcmin$ field, focusing on three galaxies: M85, NGC 4394 (east), and IC 3292 (west). Panel (a) displays the $g$-band image taken with CFHT/MegaCam (Paper I), showing fine structures of M85 such as shells and ripples, reaching galaxies on either side. Especially, prominent shells are extended along the NE-SW direction. In addition, we detect a warped faint stellar halo of IC 3292. The spatial distribution of the BGCs is more extended than that of the RGCs. Interestingly, the GGCs are lined up in the direction of NE-SW. We will investigate and discuss the spatial and kinematic peculiarities of the GC subpopulations in the following sections.

	\begin{figure}[hbt]
\epsscale{1}
\includegraphics[width=\columnwidth]{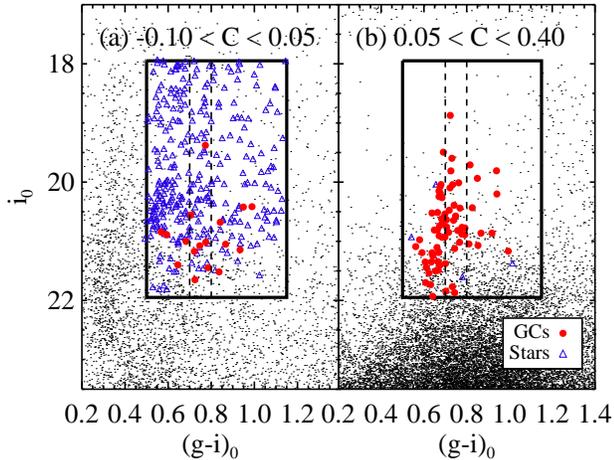}
\caption{$i_0 -(g-i)_0$ CMDs for the point-like sources (dots) from Paper I with (a) $-0.10 < C < 0.05$ (point sources) and (b) $0.05 < C < 0.40$ (slightly extended sources) in the survey region.
The blue open triangles and red filled circles represent the foreground stars and GCs confirmed in this study, respectively.
The large box shows the color and magnitude criteria used for the target selection.
The two vertical dashed lines at $(g-i)_0$ = 0.7 and 0.8 mark the color criteria to divide GCs into blue, green, and red.
\label{fig:target_cmd}}
	\end{figure}
	
{Figure \ref{fig:target_cmd}} shows $i_0-(g-i)_0$ color magnitude diagrams for point-like sources (gray dots) detected in the CFHT/MegaCam images (Paper I) as well as the GCs (red filled squares) and foreground stars (open diamonds) confirmed in this study.
We divided the objects into two groups according to their inverse concentration indices.
The inverse concentration index $C$ is defined as the difference between 4- and 8-pixel-diameter $i$-band aperture-corrected magnitudes. This parameter is broadly used to distinguish slightly extended sources from point-like sources \citep[e.g.][Paper I]{dur14}.
We found that 81\% of the confirmed GCs have $C$ values higher than 0.05, while 98\% of the confirmed foreground stars have lower $C$ values.
This indicates that most GCs in M85 are slightly extended in the CFHT/MegaCam images.

	\begin{figure}[t]
\epsscale{1}
\includegraphics[width=\columnwidth]{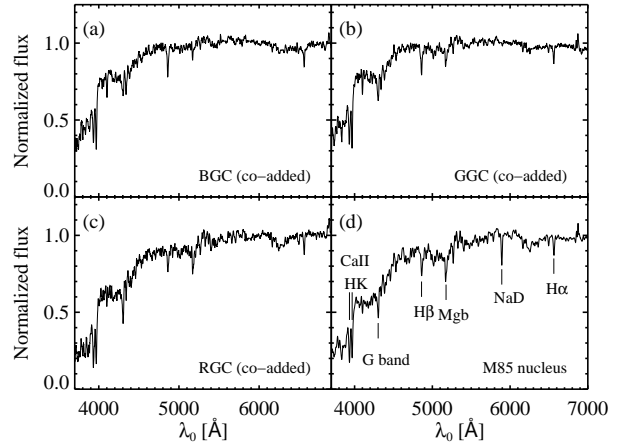}
\caption{Co-added spectra of (a) 41 BGCs, (b) 32 GGCs, and (c) 16 RGCs. Panel (d) shows the flux-calibrated spectrum of the M85 nucleus where prominent absorption lines are marked.
All spectra are plotted in the rest frame, smoothed using a boxcar filter with a size of 10 \rm{\AA}, and normalized at 5950-6050 \rm{\AA}.
\label{fig:sample_spec}}
	\end{figure}

	\subsection{Stellar population of GC subpopulations}


We estimated mean ages and metallicities of the GC subpopulations using their co-added spectra to compare their formation epochs.
{Figure \ref{fig:sample_spec}} shows the co-added spectra of three GC subpopulations (BGCs, GGCs, and RGCs) as well as the spectrum of the M85 nucleus.
The co-added spectra have higher signal-to-noise ratios ranging from 28 to 45, compared to the individual target spectra.
Several prominent absorption lines are marked in panel (d), which are identified in the other spectra as well.
The spectrum of the M85 nucleus shows broader absorption lines than those of the GCs because of larger velocity dispersion of the M85 nucleus.

We measured the Lick indices using the EZ\_Ages package \citep{gs08}.
The Lick indices have been widely used to measure ages and metallicities of old simple stellar populations \citep{bur84,wor94,wo97,tra98}. 
The stacked spectra were smoothed with Lick resolution ($\sim$ 9 \rm{\AA}).
We adopted a $\chi^2$ minimization technique using the residual between the observed Lick indices and the model prediction values to determine ages and metallicities of the GCs \citep{pfb04}.
{We used flux-calibrated stellar population models of Lick indices from \citep{tmj11} of which the ages range from 0.1 to 15 Gyr, the metallicities [Z/H] from --2.25 to +0.67, and the $\alpha$-element abundances [$\alpha$/Fe] from --0.3 to +0.5.
In the beginning, we used all Lick indices except CN1, CN2, Ca4227, and NaD indices for the fitting. The CN1, CN2, and Ca4227 indices are too sensitive to nitrogen abundances that are not well calibrated in the models we adopted, and the NaD index strongly depends on the amount of interstellar absorption. Afterwards, we calculated the $\chi^2$ values with the rest of Lick indices, using iterative 2$\sigma$ clipping.} The detailed process is described in Paper II.

We estimated uncertainties of the mean ages, [Z/H], and [$\alpha$/Fe] using a bootstrapping method.
For each subpopulation, we randomly chose the same number of GCs from the actual parent data allowing replacement, stacked their spectra, and measured ages and metallicities with the spectra. 
After repeating this procedure 1000 times, we identified 16th and 84th percentiles from the results.
The differences between these values and the results derived from the actual data were adopted as the uncertainties.

\begin{deluxetable}{l c c c c}
\tablecaption{Mean ages, [Z/H], and [$\alpha$/Fe] values of the M85 nucleus and GC subpopulations \label{tab:sp}}
\tablewidth{0pt}
\tablehead{
\colhead{Target} & \colhead{Age} & \colhead{[Z/H]} & \colhead{[$\alpha$/Fe]} & \colhead{S/N}\\
\colhead{} & \colhead{(Gyr)} & \colhead{(dex)} & \colhead{(dex)} & \colhead{at 5000 \rm{\AA}}
}
\startdata
BGC & $14.0^{+<0.1}_{-5.5}$ & $-1.49^{+0.28}_{-0.04}$ & $0.24^{+0.13}_{-0.37}$ & 38 \\
GGC & $11.1^{+3.4}_{-3.2}$ & $-0.91^{+0.20}_{-0.12}$ & $0.15^{+0.09}_{-0.09}$ & 45 \\
RGC & $9.9^{+2.1}_{-3.7}$ & $-0.45^{+0.22}_{-0.11}$ & $0.19^{+0.06}_{-0.16}$ & 28 \\
M85 nucleus & $2.9^{+<0.1}_{-<0.1}$ & $0.26^{+<0.01}_{-0.01}$ & $0.24^{+<0.01}_{-0.01}$ & 304 \\
\enddata
\end{deluxetable}

{Table \ref{tab:sp}} lists the mean ages, [Z/H], and [$\alpha$/Fe] values of the GC subpopulations and the nucleus of M85. 
The mean ages of BGCs, GGCs, and RGCs are 14.0$^{+<0.1}_{-5.5}$, 11.1$^{+3.4}_{-3.2}$, and 9.9$^{+2.1}_{-3.7}$ Gyr, respectively, which are consistent within uncertainties.
On the other hand, the GC subpopulations show differences in their metallicities. 
The BGCs and RGCs have mean metallicities of [Z/H] = $-1.49^{+0.28}_{-0.04}$ and $-0.45^{+0.22}_{-0.11}$, respectively, showing that they are the most metal-poor and the most metal-rich populations.
{They are consistent with those in other galaxies with the same luminosity of M85 \citep[$M_B = -21.28$ mag;][]{bst85}, according to the GC mean metallicity and host galaxy luminosity \citep{pen06}.}
The GGCs in M85 have a mean metallicity of [Z/H] = $-0.91^{+0.20}_{-0.12}$, which is between those of the BGCs and RGCs.
The $\alpha$-element abundances of the GC subpopulations range from 0.15 to 0.24, but it is hard to tell any significant difference because of large uncertainties.

We derived the age and metallicity of the M85 nucleus to be about 2.9 Gyr and [Z/H] = +0.26, which are similar to those derived using the Gemini/GMOS optical spectrum in Paper II.
This shows that the stellar population in the M85 nucleus is much younger and more metal-rich than any GC subpopulation identified in this study.
We could not find any GC population that were formed when the central star formation occurred in the nucleus of M85, while Paper II found an intermediate-age GC population within $R = 3\arcmin$.
This implies either that the intermediate-age GCs rarely exist in the outer region of M85.

	\subsection{Kinematics of the GC System in M85}

\begin{deluxetable*}{c c c c c c c c c}
\setlength{\tabcolsep}{0.15in}
\tablecaption{Kinematics of GC Systems in M85 \label{tab:kinematics}}
\tablewidth{0pt}
\tablehead{
\colhead{$R$} & \colhead{$\overline{R}$} & \colhead{N} & \colhead{$\overline{v_{\rm r}}$} & \colhead{$\sigma_{\rm r}$} & \colhead{$\Omega R$} & \colhead{$\Theta_0$} & \colhead{$\sigma_{\rm r, cor}$} & \colhead{$\Omega R / \sigma_{\rm r(,cor)}$} \\
\colhead{(arcmin)} & \colhead{(arcmin)} & \colhead{} & \colhead{(\kmsend)} & \colhead{(\kmsend)} & \colhead{(\kmsend)} & \colhead{(deg)} & \colhead{(\kmsend)}
}
\startdata
\multicolumn{9}{c}{All GCs: 89 GCs with $0.6 < (g-i)_0 < 1.0$}\\
\hline
 1.5--31.0 & 8.6 & 89 & 754$^{+19}_{-19}$ & 165$^{+9}_{-14}$ & 80$^{+52}_{-12}$ & 156$^{+26}_{-46}$ & 163$^{+11}_{-11}$ & 0.49$^{+0.33}_{-0.07}$ \\
 1.5--5.9 & 3.9 & 48 & 740$^{+23}_{-25}$ & 152$^{+10}_{-17}$ & 150$^{+35}_{-37}$ & 188$^{+7}_{-13}$ & 132$^{+10}_{-17}$ & 1.13$^{+0.33}_{-0.27}$ \\
 6.0--31.0 & 14.1 & 41 & 769$^{+30}_{-30}$ & 177$^{+17}_{-25}$ & 198$^{+100}_{-78}$ & 78$^{+45}_{-13}$ & $\cdots$ & 1.12$^{+0.59}_{-0.43}$ \\
\hline
\multicolumn{9}{c}{BGCs: 41 GCs with $0.6 < (g-i)_0 < 0.7$}\\
\hline
 1.5--23.9 & 9.4 & 41 & 727$^{+30}_{-29}$ & 168$^{+14}_{-23}$ & 50$^{+68}_{-64}$ & 162$^{+22}_{-116}$ & $\cdots$ & 0.30$^{+0.38}_{-0.28}$ \\
 1.5--5.9 & 4.4 & 17 & 706$^{+39}_{-40}$ & 144$^{+18}_{-34}$ & 59$^{+86}_{-95}$ & 44$^{+98}_{-11}$ & $\cdots$ & 0.41$^{+0.64}_{-0.71}$ \\
 6.0--23.9 & 12.9 & 24 & 742$^{+38}_{-40}$ & 181$^{+17}_{-31}$ & 136$^{+64}_{-66}$ & 175$^{+19}_{-49}$ & $\cdots$ & 0.75$^{+0.32}_{-0.27}$ \\
\hline
\multicolumn{9}{c}{GGCs: 32 GCs with $0.7 < (g-i)_0 < 0.8$}\\
\hline
 1.9--31.0 & 9.5 & 32 & 812$^{+30}_{-27}$ & 156$^{+15}_{-24}$ & $\cdots$ & $\cdots$ & $\cdots$ & $\cdots$ \\
 1.9--5.9 & 3.6 & 16 & 818$^{+39}_{-34}$ & 139$^{+13}_{-28}$ & $\cdots$ & $\cdots$ & $\cdots$ & $\cdots$ \\
 6.6--31.0 & 15.5 & 16 & 805$^{+48}_{-44}$ & 170$^{+21}_{-45}$ & $\cdots$ & $\cdots$ & $\cdots$ & $\cdots$ \\
\hline
\multicolumn{9}{c}{RGCs: 16 GCs with $0.8 < (g-i)_0 < 1.0$}\\
\hline
 1.7--20.3 & 4.8 & 16 & 704$^{+37}_{-37}$ & 141$^{+14}_{-26}$ & 203$^{+41}_{-42}$ & 197$^{+5}_{-7}$ & 95$^{+20}_{-36}$ & 2.15$^{+1.65}_{-0.49}$ \\
 1.7--5.3 & 3.7 & 15 & 695$^{+43}_{-38}$ & 142$^{+16}_{-31}$ & 208$^{+41}_{-42}$ & 196$^{+5}_{-7}$ & 66$^{+16}_{-25}$ & 3.16$^{+2.37}_{-0.53}$ \\
\enddata
\end{deluxetable*}

We investigate kinematic properties of the GC system of M85 such as mean radial velocities, rotation properties, and velocity dispersions.
We used a numerical bootstrapping method to estimate uncertainties of all kinematic parameters.
We randomly chose the same number of GCs from the parent data allowing replacement to construct a mock data set, and derived their kinematic parameters.
After repeating this procedure 1000 times, we identified 16th and 84th percentiles for the results, which corresponds to 68\% confidence intervals.
We adopted the differences between these values and the parameters measured from the actual parent data as uncertainties.

We compare the mean radial velocities, rotation properties, and velocity dispersions of the GC subpopulations in the following sections.
{Table \ref{tab:kinematics}} lists the kinematic parameters derived for the entire GC, BGC, GGC, and RGC systems.

		\subsubsection{Mean radial velocities}

	\begin{figure}[t]
\epsscale{1}
\includegraphics[width=\columnwidth]{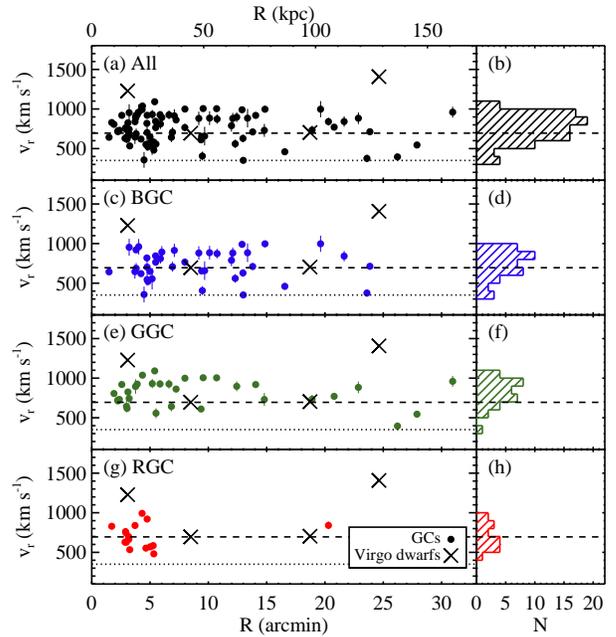}
\caption{(a) Radial velocity vs. galactocentric distance from M85 for all GCs. 
The filled circles and crosses represent the GCs and Virgo dwarf galaxies in the survey region \citep{kim14}, respectively.
The dashed and dotted lines indicate the radial velocity of M85 nucleus and the criteria used to divide the targets into GCs and foreground stars (\vrad = 350 \kmsend), respectively.
(b) Radial velocity distribution of all GCs.
Panels (c)-(d), (e)-(f), and (g)-(h) are the same as (a)-(b), but for BGCs, GGCs, and RGCs, respectively.
\label{fig:r_vr}}	
	\end{figure}	

{Figure \ref{fig:r_vr}} shows radial velocity distributions of all GCs and GC subpopulations confirmed in this study as a function of galactocentric distance from M85.
All GCs are located in the radial range of $1\farcm5 < R < 31\farcm0$ that corresponds to 8 kpc $< R <$ 162 kpc.
The BGCs and GGCs are found in the entire radial range, but all RGCs are within $R = 6\arcmin$ except for one.

The mean radial velocity of all GCs is $\overline{v_{\rm r}} =754 \pm 19$ \kmsend, which is 58 \kms higher than the radial velocity of the M85 nucleus, $696 \pm 16$ \kmsend.
About two thirds of the GCs have radial velocities higher than the nucleus velocity ({Figure \ref{fig:r_vr}(b)}).
This number excess of the high velocity GCs is mainly contributed by the GGCs that have a much higher mean radial velocity of $\overline{v_{\rm r}} = 812^{+30}_{-27}$ \kmsend.
If the GGCs are excluded, the mean radial velocity of the GCs drops to $\overline{v_{\rm r}} =721^{+23}_{-21}$ \kmsend.
In addition, the BGCs and RGCs have the mean radial velocities of $\overline{v_{\rm r}} = 727^{+30}_{-29}$ \kms and $704 \pm 37$ \kmsend, respectively.
These mean radial velocities measured without the GGC population are consistent with the radial velocity of the M85 nucleus within uncertainties. 

\begin{figure}[t]
\epsscale{1}
\includegraphics[width=\columnwidth]{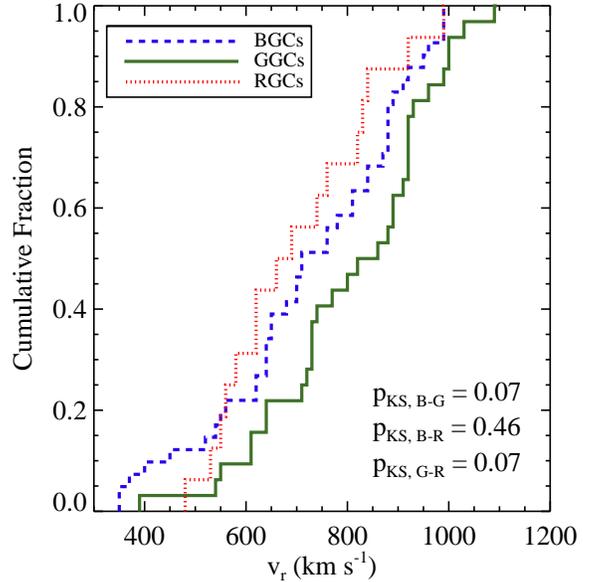}
\caption{Cumulative radial velocity distributions of BGCs (dashed line), GGCs (solid line), and RGCs (dotted line).
\label{fig:cum_vdist}}
	\end{figure}

{We performed two-sided Kolmogorov-Smirnov (K-S) tests to compare the radial velocity distributions of GC subpopulations. {Figure \ref{fig:cum_vdist}} shows the cumulative radial velocity distributions of BGCs, GGCs, and RGCs in M85. The $p$-value for the BGCs and RGCs is 0.46, from which we cannot tell a clear difference between their velocity distributions. However, the $p$-values for the B/RGCs and GGCs are 0.07, meaning that the GGCs have a radial velocity distribution clearly distinct from those of both BGCs and RGCs.}

In {Figure \ref{fig:r_vr}} we also plot the radial velocities of dwarf galaxies in the Virgo Cluster for comparison with the M85 GC kinematics.
We adopted the radial velocities of Virgo galaxies from the Extended Virgo Cluster Catalog \citep[EVCC;][]{kim14}.
There are only four dwarf galaxies in our survey region: VCC 797 (EVCC 556) at $R = 3\farcm1$, VCC 751 (EVCC 529) at $R = 8\farcm5$, EVCC 671 at $R = 18\farcm7$, and EVCC 629 at $R = 24\farcm6$.
Two of them have radial velocities similar to the M85 velocity (\vrad = 696 \kms for VCC 751 and 703 \kms for EVCC 617), but the other two have much higher velocities (\vrad = 1228 \kms for VCC 797 and 1408 \kms for EVCC 629).
Because of the small number statistics, it is not clear whether these dwarf galaxies constitute a distinguishable group associated with M85 or they are governed by the gravitational potential of the Virgo Cluster.
We will discuss the kinematic differences between M85 GCs and Virgo dwarf galaxies with regard to the dark matter extent of M85 in Section 4.2.

		\subsubsection{Rotation Properties}

	\begin{figure*}[t]
\epsscale{1}
\includegraphics[width=\textwidth]{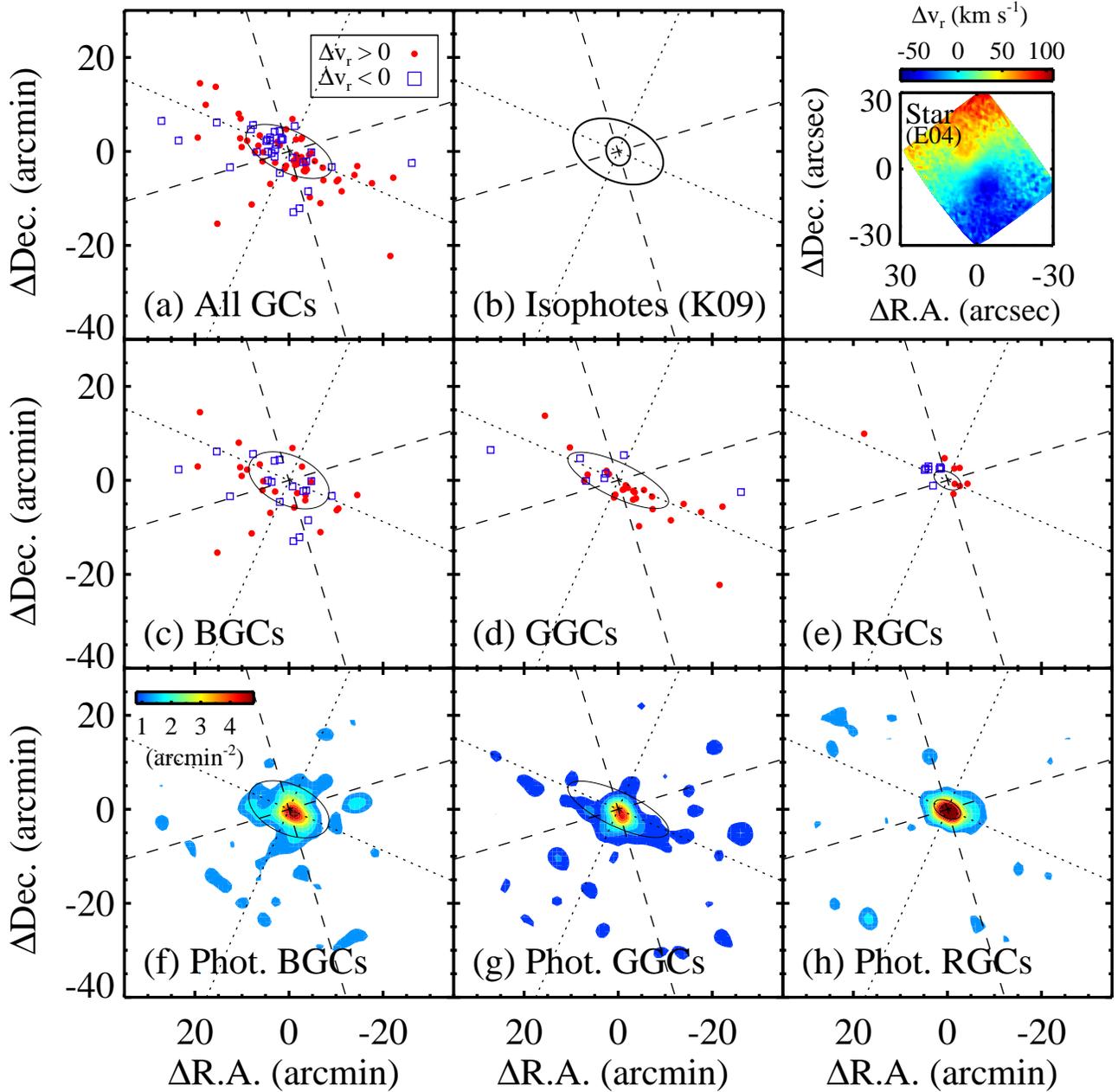}
\caption{(a) Spatial distribution of all GCs confirmed in this study. The red filled circles and blue open squares represent the GCs with the radial velocities higher and lower than that of the M85 nucleus, respectively.
The dashed and dotted lines represent the photometric major/minor axes of the isophotes derived at $r_{\rm maj}=3\arcmin$ and 10$\arcmin$, respectively \citep{kor09}.
(b) Stellar light isophotes at $r_{\rm maj}=3\arcmin$ and 10$\arcmin$ \citep{kor09}.
(c)-(e) Same as (a), but for BGCs, GGCs, and RGCs. {The solid-line ellipses indicate the dispersion ellipses derived with the GC subgroups within $R = 20\arcmin$.}
(f)-(h) Smoothed surface number density map of the parent photometric sample of BGCs, GGCs, and RGCs, color-coded by the number density.
{The solid-line ellipses are the same as in panel (c)-(e).}
For comparison, the SAURON stellar velocity field for the $30\arcsec \times 30\arcsec$ region is given in the upper right corner \citep{ems04}. 
\label{fig:v_spatial}}
	\end{figure*}

{Figure \ref{fig:v_spatial}(a)} shows the spatial distribution of the GCs along with their radial velocities.
The GCs are strongly concentrated around the galaxy center and show an elongated spatial distribution.
For comparison, we plot the isophotes of the M85 stellar light in {Figure \ref{fig:v_spatial}(b)}.
The position angles and ellipticities of the isophotes change from 16$^{\circ}$ to 66$^{\circ}$ and from 0.18 to 0.38 as the semi-major axis $R_{\rm maj}$ increases from 3$\arcmin$ to 10$\arcmin$ \citep{kor09}.
The GCs show a spatial distribution elongated along the major axis of the isophote of M85 at $R_{\rm maj} = 10\arcmin$.

We also investigate the spatial distributions of the GC subpopulations separately ({Figure \ref{fig:v_spatial}(c)-(e)}).
The BGCs are sparsely distributed without any trend in their radial velocities.
In contrast, the RGCs are strongly concentrated around the galaxy center and show a clear spatial segregation between high and low radial velocity GCs.
We consider that this spatial segregation of the RGCs indicates a rotation signature of the RGC system.
Interestingly, the GGCs are strongly aligned along the photometric major axis of the outer stellar isophote of M85, especially for the ones with higher relative velocities.
These features in the spatial distributions of the spectroscopic samples are similar to those of the photometric samples with $18 < i_0 < 22$ identified by Paper I ({Figure \ref{fig:v_spatial}(f)-(h)}).

\begin{figure}[hbt]
\epsscale{1}
\includegraphics[width=\columnwidth]{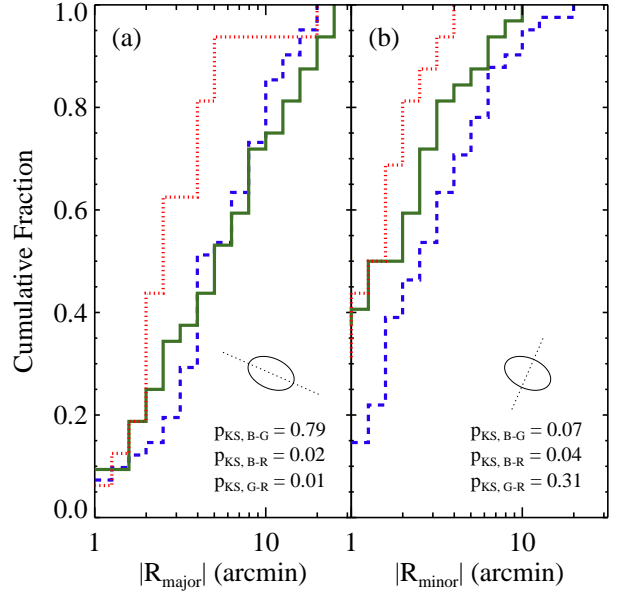}
\caption{(a) Cumulative major axis distance distributions of BGCs (dashed line), GGCs (solid line), and RGCs (dotted line). (b) Same as (a), but for minor axis distance.
\label{fig:mjmn_dist}}
	\end{figure}

{For quantitative comparison between spatial distributions of the GC subgroups, we performed two-sided K-S tests on their major and minor axis distances, adopting the major and minor axes of the isophote of M85 at $R_{\rm maj} = 10\arcmin$. {Figure \ref{fig:mjmn_dist}} shows the cumulative major and minor axis distances of the GC subpopulations. The BGCs and GGCs are more extended along the major axis than the RGCs. The RGCs have a major axis distance distribution clearly distinct from the other two ($p$ = 0.01-0.02), while the BGCs and GGCs have similar major axis distribution ($p$ = 0.79). On the other hand, along the minor axis, the BGCs are the most extended among the GC subgroups, having $p$-values of 0.07 and 0.04 for the comparison with GGCs and RGCs, respectively. The GGCs and RGCs could not be clearly distinguished with the $p$-value of 0.31.}

{In addition, we measured shape parameters such as ellipticity and position angle of the GC systems, assuming that the GCs constitute an ellipse. We used the dispersion ellipse of the bivariate normal frequency function of position vectors \citep{tw53}. The dispersion ellipse represents the contour at which the density is 0.61 times the maximum density of a set of points. This method has been often used for quantitative analysis on the two-dimensional distributions of galaxies in a galaxy cluster \citep{cm80, bur04, hl07} or GCs in a galaxy \citep{mhh94, har11, pl13}. We followed their analysis and derived the parameter uncertainties from 16th and 84th percentiles from the bootstrapping procedure with 1000 trials. We calculated the following five moments,
\begin{subequations}
	\begin{align}
\begin{split}
    \mu_{10} = \frac{1}{N}\sum_{i=1}^{N} x_i ,
\end{split}\\
\begin{split}
    \mu_{01} = \frac{1}{N}\sum_{i=1}^{N} y_i, 
\end{split}\\
\begin{split}
	\mu_{20} = \frac{1}{N}\sum_{i=1}^{N} x_i^2 - \bigg(\frac{1}{N}\sum_{i=1}^{N} x_i\bigg)^2,
\end{split}\\
\begin{split}
	\mu_{02} = \frac{1}{N}\sum_{i=1}^{N} y_i^2 - \bigg(\frac{1}{N}\sum_{i=1}^{N} y_i\bigg)^2,
\end{split}\\
\begin{split}
	\mu_{11} = \frac{1}{N}\sum_{i=1}^{N} x_iy_i - \frac{1}{N^2}\sum_{i=1}^{N} x_i \sum_{i=1}^{N} y_i.
\end{split}
	\end{align}
\end{subequations}
Using these moments, the semimajor and semiminor axes of the dispersion ellipse, $\Gamma_A$ and $\Gamma_B$, are derived with the following equation,
\begin{equation}
\begin{vmatrix}
\mu_{20}-\Gamma^2 & \mu_{11}\\
\mu_{11} & \mu_{02}-\Gamma^2
\end{vmatrix}
=0.
\end{equation}
The position angle of the major axis is given by
\begin{equation}
\theta = \mathrm{cot}^{-1}\bigg(-\frac{\mu_{02}-\Gamma_{A}^2}{\mu_{11}}\bigg)+\frac{\pi}{2},
\end{equation} 
and the ellipticity is
\begin{equation}
\epsilon = 1 - \frac{\Gamma_B}{\Gamma_A}.
\end{equation}

{Figure \ref{fig:v_spatial}(a)} shows the dispersion ellipse for the entire GC sample confirmed in this study. Its position angle and ellipticity are $\theta = 65^{+5}_{-4}\arcdeg$ and $e = 0.55 \pm 0.06$. For comparison, we also derived the dispersion ellipses for GC subgroups, as shown in {Figure \ref{fig:v_spatial}(c)-(e)}. For the RGC system, we excluded a RGC with $R > 20\arcmin$ that is an outlier of the overall distribution of the RGCs.
The position angles of the dispersion ellipses for BGCs, GGCs, and RGCs are $\theta = 65\arcdeg \pm 8\arcdeg, 65\arcdeg \pm 6\arcdeg,$ and $64^{+12}_{-9}\arcdeg$, respectively. They are consistent with each other, and similar to the photometric position angle of the isophote of M85 at $R_{\rm maj} = 10\arcmin$ ($\sim 66\arcdeg$). On the other hand, the ellipticity of the GGC system ($e = 0.69 \pm 0.04$) is 2$\sigma$ higher than those of both BGC and RGC systems ($e = 0.42^{+0.11}_{-0.10}$ and $0.44^{+0.12}_{-0.13}$), which means that the spatial distribution of the GGCs is more elongated than those of the BGCs and RGCs.}

	\begin{figure*}[t]
\epsscale{1}
\includegraphics[width=\textwidth]{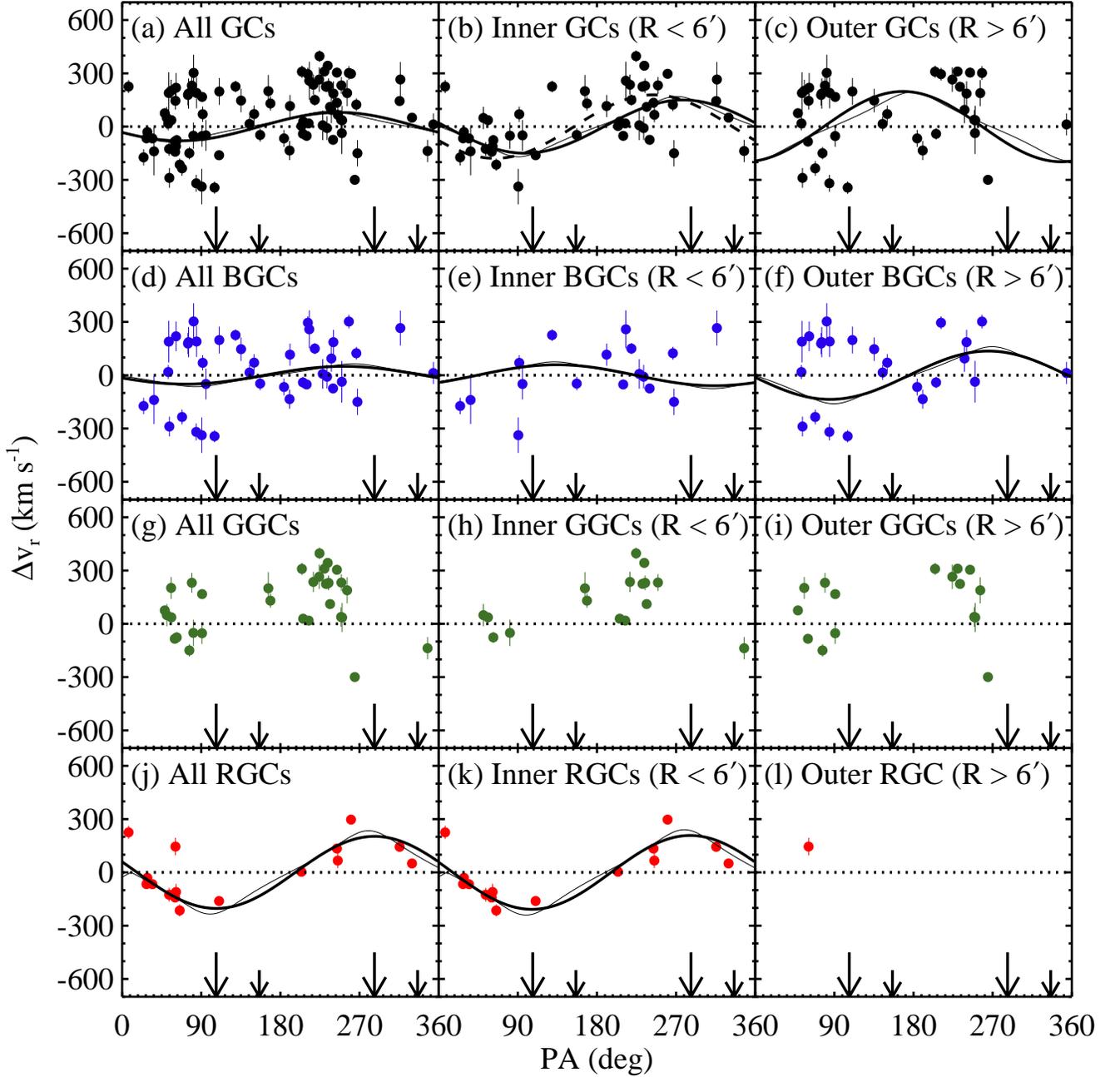}
\caption{Radial velocities relative to the systemic velocity as a function of position angle for (a) all GCs, (b) inner GCs ($R < 6\arcmin$), and (c) outer GCs ($R > 6\arcmin$).
Panels (d)-(f), (g)-(i), and (j)-(l) are the same as panels (a)-(c), but for BGCs, GGCs, and RGCs, respectively.
The thick solid and dotted lines indicate the best fit rotation curves and zero velocity line, respectively. {The thin solid lines show the rotation curves derived with the rotation axis ratio of $q = 0.625$ (see texts).}
The dashed line in panel (b) shows the best fit rotation curve derived with 20 M85 GCs within $R = 3\arcmin$ (Paper II).
The large and small vertical arrows mark the photometric minor axis of M85 at $r_{\rm maj}=3\arcmin$ and 10$\arcmin$, respectively \citep{kor09}.
\label{fig:pa_vr1}}
	\end{figure*}

In addition to the differences in the spatial distributions of the GC subsystem, we examine their differences in the rotation features.
{Figure \ref{fig:pa_vr1}} shows the radial velocities of GCs as a function of position angle with the best-fit rotation curves.
The GCs rotating along a given axis in the plane of the sky have radial velocities as a function of sinusoidal position angle.
We measured the rotation amplitude and position angle of the rotation axis for the GC system by fitting the data with the following function:
\begin{equation}
    v_{\rm r}(\Theta)~=~v_{\rm sys} + (\Omega R)~{\rm sin}(\Theta-\Theta_0),
\end{equation}
where the $v_{\rm sys}$ is the systemic velocity, $\Omega R$ is the rotation amplitude, and $\Theta_0$ is the orientation of the rotation axis.
We assumed the systemic velocity to be the radial velocity of the M85 nucleus, $v_{\rm sys}$ = 696 \kmsend, as derived in this study.

We fitted the data with this function for the entire GC system of M85 ({Figure \ref{fig:pa_vr1}(a)}).
The rotation amplitude and the orientation of the rotation axis of the GC system are $\Omega R = 80^{+52}_{-12}$ \kms and $\Theta_0 = 156^{+26}_{-46}\arcdeg$.
The rotation axis is close to the minor axis of the stellar isophote at $R_{\rm maj} = 10\arcmin$.
We calculated a rotation parameter, $\Omega R/\sigma_{\rm r,cor}$, defined as the ratio between the rotation amplitude and the rotation-corrected velocity dispersion.
The entire GC system has the rotation parameter of $\Omega R/\sigma_{\rm r,cor} = 0.49^{+0.33}_{-0.07}$ with the rotation-corrected velocity dispersion of $\sigma_{\rm r, cor} = 163 \pm 11$ \kmsend.
This rotation parameter value is consistent with those of GC systems of massive early-type galaxies with luminosity similar to M85 \citep[e.g. $0.45^{+0.25}_{-0.24}$ for M84 and $0.65^{+0.27}_{-0.22}$ for M60;][]{ala16, hwa08}.

We found that the BGC and RGC systems show rotation properties significantly different from each other ({Figure \ref{fig:pa_vr1}(d) and (j)}).
The RGC system shows a strikingly strong rotation feature with a rotation amplitude of $\Omega R = 203^{+41}_{-42}$ \kmsend, which is almost a disk-like rotation.
On the other hand, the rotation amplitude of the BGC system is close to zero with a large uncertainty ($\Omega R = 50^{+68}_{-64}$ \kmsend).
The orientation of the rotation axis of the BGC system has a large uncertainty because its rotation feature is negligible ($\Theta_0 = 162^{+22}_{-116} \arcdeg$), but for the RGC system, it is precisely measured with a small uncertainty ($\Theta_0 = 197^{+5}_{-7} \arcdeg$).
The rotation parameters of the BGC and RGC systems are $\Omega R/\sigma_{\rm r} = 0.30^{+0.38}_{-0.28}$ and $\Omega R/\sigma_{\rm r, cor} = 2.15^{+1.65}_{-0.49}$, respectively. We did not apply the rotation correction to the velocity dispersion for the BGC system because of its negligible rotation feature.

We did not derive the rotation parameters for the GGC system because most of GGCs have radial velocities higher than the systemic velocity ($\Delta v > 0$ \kmsend) and are concentrated only at the position angle of 70$\arcdeg$ and 240$\arcdeg$ ({Figure \ref{fig:pa_vr1}(g)}).

We additionally divide the entire GC, BGC, GGC, and RGC samples into two groups, inner and outer systems, with a radial criterion of $R = 6\arcmin$, and investigate their kinematics.
The inner GC system has a rotation amplitude of $\Omega R = 150^{+35}_{-37}$ \kms and a rotation axis of $\Theta_0 = 188^{+7}_{-13}\arcdeg$, which are marginally consistent with those derived from the small GC sample within $R = 3\arcmin$ ({Figure \ref{fig:pa_vr1}(b)}; Paper II). 
The rotation-corrected velocity dispersion for the inner GC system is $\sigma_{\rm r, cor} = 132^{+10}_{-17}$ \kmsend, resulting in the rotation parameter of $\Omega R/\sigma_{\rm r, cor} = 1.13^{+0.33}_{-0.27}$. 
This rotation parameter value is two times higher than that derived for the entire GC sample.
This strong rotation of the inner GC system is mainly contributed by the RGCs.
All RGCs are located within $R = 6\arcmin$ except for one, rotating strongly with a rotation amplitude of $\Omega R = 208^{+41}_{-42}$ \kms and a rotation parameter of $\Omega R/\sigma_{\rm r, cor} = 3.16^{+2.37}_{-0.53}$ ({Figure \ref{fig:pa_vr1}(k)}).
On the other hand, the inner BGC system does not show any significant rotation features ({Figure \ref{fig:pa_vr1}(e)}).
The inner BGC system has a rotation parameter of $\Omega R/\sigma_{\rm r} = 0.41^{+0.64}_{-0.71}$, which is similar to the entire BGC system.
Most of the inner GGCs are concentrated at the position angle of 240$\arcdeg$, indicating a bulk motion ({Figure \ref{fig:pa_vr1}(h)}).

The outer GC system has a rotation amplitude of $\Omega R = 198^{+100}_{-78}$ \kmsend, which is comparable with the velocity dispersion of $\sigma_{\rm r} = 177^{+17}_{-25}$ \kmsend.
The rotation axis of the outer GC system is $\Theta_0 = 78^{+45}_{-13} \arcdeg$, which is totally different from that of the inner GC system ({Figure \ref{fig:pa_vr1}(c)}).
This is because the fitting results are sensitive to the outer GGCs only concentrated at the position angle of 70$\arcdeg$ and 240$\arcdeg$ ({Figure \ref{fig:pa_vr1}(i)}).
If we only consider the outer BGC system, the rotation amplitude and orientation of the rotation axis are $\Omega R = 136^{+64}_{-66}$ \kms and $\Theta_0 = 175^{+19}_{-49} \arcdeg$ ({Figure \ref{fig:pa_vr1}(f)}).
The rotation parameter of the outer BGC system ($\Omega R/\sigma_{\rm r} = 0.75^{+0.32}_{-0.27}$) is higher than that of the inner BGC system, but consistent within uncertainties.

{Additionally, we checked a possibility that there are any significant changes in the kinematic parameter measurements if the spatial elongations of the GC systems are considered in the fitting. \citet{pro09} suggested a rotation model considering a rotation axis ratio:
\begin{equation}
	v_{\rm mod}~=~v_{\rm sys} \pm \frac{v_{\rm rot}}{\sqrt{1+\Big( \frac{\displaystyle{\rm tan}(\Theta - {\rm PA_{kin}})}{\displaystyle q}} \Big) ^2}
\end{equation}
where PA$_{\rm kin}$ is the kinematic position angle defined as the angle from the north to the maximum receding part of the velocity map, and $q$ is the rotation axis ratio. The kinematic position angle PA$_{\rm kin}$ is, by definition, different from the rotation axis orientation $\Theta_0$ in equation (5) by 90$^\circ$.
This equation corresponds to the sinusoidal function we used if the axis ratio $q$ is equal to 1. 

We adopted the photometric axis ratio of the isophotes at $R_{\rm maj} = 10\arcmin$ ($q$ = 0.625), which is the most extreme case for the elongation. The fitting results are also plotted in {Figure \ref{fig:pa_vr1}} for comparison. 
The rotation amplitudes derived with the axis ratio of $q$ = 0.625 is slightly higher than the previous measurements. However, the mean difference is only 18 \kmsend, which is much smaller than the measurement uncertainties. The orientations do not show any significant differences as well. Therefore, we conclude that our kinematic parameter measurements are irrelevant to the elongated spatial distributions of the GC systems.}

		\subsubsection{Mean Radial Velocity And Velocity Dispersion Profiles}

	\begin{figure}[hbt]
\epsscale{1}
\includegraphics[width=\columnwidth]{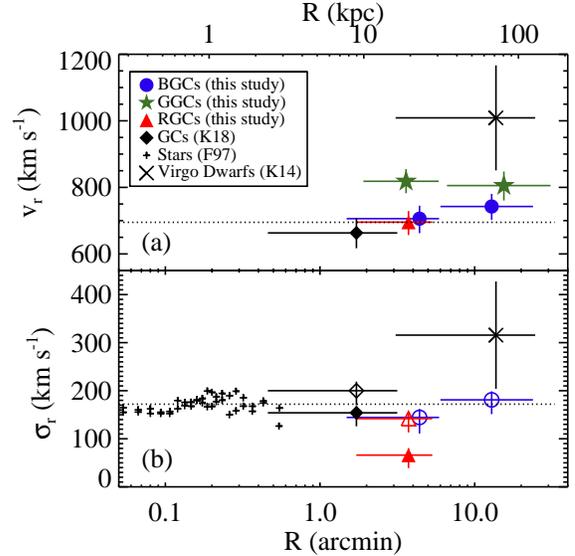}
\caption{(a) Mean radial velocity profiles for the M85 GCs at $R < 2\farcm5$ (diamonds; Paper II), M85 BGCs/GGCs/RGCs confirmed in this study (circles/stars/triangles), and Virgo dwarfs in the survey region \citep[cross;][]{kim14}. The horizontal dotted line indicates the radial velocity of the M85 nucleus measured in this study.
(b) Same as (a), but for radial velocity dispersion profiles. The pluses represent the stars in the central region of M85 \citep{fis97}. The horizontal dotted line indicates the velocity dispersion of the central region of M85 measured from the integrated stellar light spectrum within $R = 1\arcsec$ (86.8 pc) (Paper II).
The open and filled symbols for the M85 GCs represent the velocity dispersions about the systemic velocity ($\sigma_{\rm r}$) and the rotation-correction velocity dispersions ($\sigma_{\rm r, cor}$), respectively.
\label{fig:rad_vd}}	
	\end{figure}

{Figure \ref{fig:rad_vd}(a)} shows the mean radial velocities of GCs and Virgo dwarf galaxies in the survey region as a function of galactocentric distance from M85. We confirmed that the GGCs have mean radial velocities higher than the other GC subpopulations regardless of distance, as shown in Section 3.3.1. The mean radial velocity of four Virgo dwarf galaxies in the survey region is much higher than that of GCs.

In {Figure \ref{fig:rad_vd}(b)}, we display velocity dispersion profiles of the GCs confirmed in this study at $1\farcm5 < R < 23\farcm9$, the GCs studied in the previous study (Paper II), and the central stars at $R < 0\farcm6$ \citep{fis97}.
The velocity dispersion of the inner BGCs is similar to that of the central stars.
The outer BGCs with mean galactocentric distance of $R = 12\farcm9$ have about 40 \kms higher velocity dispersion than the inner BGCs, but this difference is within uncertainties.
The RGCs tightly follow the fitted rotation curve ({Figure \ref{fig:pa_vr1}(j)}), resulting in a dramatic change of the velocity dispersion after the rotation correction.
The rotation-corrected velocity dispersion of the RGCs at $R = 3\farcm7$ ($\sigma_{\rm r, cor} = 66^{+16}_{-25}$ \kmsend) is much lower than those of both the central stars and the BGCs.
This indicates that the BGCs and RGCs trace different halo components of M85 (see Section 4.2).

\section{Discussion}

	\subsection{Peculiar Motions of the M85 GC System}

We confirm a strong rotation of the inner GC system of M85 in this study.
This rotation was previously discovered from a small sample of 20 GCs in the inner region by Paper II.
This study based on the four times larger sample found that this rotation is mainly due to the RGC system.
The kinematic differences between the BGC and RGC systems have been often found in massive early-type galaxies \citep[e.g.][]{lee10, pot13}.
In general, the RGCs show a tighter correlation in the rotation velocity and velocity dispersion with those of the underlying stars of their host galaxies than the BGCs.
This indicates that the RGCs were formed when the bulk of stars in their host galaxies were formed, but the BGCs have different origins such as accretion from low-mass galaxies.

In the case of M85, the most interesting points found in this study are that 1) the RGC system strongly rotates and 2) its rotation feature does not even correspond to that of the central stars.
The rotation feature of the stars in the central region ($R < 20\arcsec$) of M85 was derived in detail by ATLAS$^{\rm 3D}$ \citep[][see also an early measurement in \citet{fis97}]{cap11, kra11, ems11}.
The kinematic position angle of the central stars is PA$_{\rm kin}$ = 19$\rlap{.}{^\circ}$5 $\pm$ 4$\rlap{.}{^\circ}$8 \citep{kra11}.
We derive the kinematic position angle of the RGC system to be 287$\arcdeg$, which is almost perpendicular to that of the central stars.
It indicates that the RGCs and central stars in M85 have undergone different experiences in their formation history.
However, the stellar kinematics is derived only in the innermost region within $R \sim 20\arcsec$, while the RGCs are located in the outer region ($1\farcm7 < R < 20\farcm3$).
Therefore, for a fair comparison, it is needed to confirm the kinematic differences between the stars and the RGC system with a stellar kinematics study of the outskirts of M85, which can be traced by planetary nebulae, for example.

In addition, we found that the GGCs constitute a stream aligned along the major axis of M85 out to $R = 31\arcmin$.
The mean radial velocity of these GGCs, \vrad = 812$^{+30}_{-27}$ \kmsend, is about 3$\sigma$ higher than that of the other GC populations.
This indicates that the GGCs may be a population infalling toward or outgoing from the M85 plane, which is associated to any previous merger or accretion event.

There have been several observational and simulation studies on GC streams in galaxies, suggesting that the GC streams are associated with stellar streams stripped from disrupting galaxies.
For example, \citet{fos14} investigated the kinematics of GCs in the Umbrella Galaxy, NGC 4651, and showed that some GCs in the faint stellar substructures are remnants produced by a minor merger event with 1:50 stellar mass ratio by comparing their kinematics with simulation data.
\citet{mac19} presented the kinematics of GCs in the outer halo of M31 and found that the GCs associated with the stellar halo substructures rotate with perpendicular orientation with respect to the GCs in the smooth halo. They interpreted that these two distinct GC populations are considered to show the signatures from two different major accretion events.
Recently, \citet{ala20} studied the GCs in the spiral galaxy NGC 5907 lying in the stellar stream. They estimated a stellar mass of the disrupted galaxy that remained the stream, using the mean metallicities of those GCs.
In addition, \citet{hug19} examined simulated galaxies and their GC systems from the E-MOSAICS project to understand the relation between physical properties of GCs in stellar streams and their host progenitors.

Likewise, we try to seek any stellar substructure in M85 associated with the GGC stream to understand the origin of the GGCs. We could not detect any faint stellar streams along the direction where the GGCs are tightly aligned. Nevertheless, we found several shell structures that are perpendicular to that direction ({Figure \ref{fig:image_bgr}(c)}). This shows a possibility that the GGCs are associated with any disrupted galaxies during minor merger events.

{If the GGCs are an accreted population that originate from a single galaxy, we can infer the stellar mass of the progenitor from the mean metallicity of the GGCs. \citet{pen06} presented a relation between mean metallicity of GCs and stellar mass of their host galaxy: [Fe/H] = ($-5.250~\pm~0.156$) + ($0.409~\pm~0.014$) log ($M_*$). This iron abundance is calibrated to the metallicity scale of \citet{zw84}, which corresponds to the total metallicity we derived in this study \citep{tmb03}. The GGCs have the mean metallicity of [Z/H] $\sim -0.91$. Using the above relation, a stellar mass of the disrupted galaxy that possibly had hosted the GGCs is expected to be $4 \times 10^{10} M_{\odot}$.}

	\subsection{Dark Matter Halo of M85}

We found that the BGCs and RGCs show different behavior in their rotation-corrected velocity profiles ({Figure \ref{fig:rad_vd}}).
The velocity dispersion profile of the BGCs is approximately flat out to $R = 12\farcm9$ (67 kpc), while the rotation-corrected velocity dispersion of the RGCs is much lower than that of the BGCs.

\citet{pl13} suggested that early-type galaxies have dual halos, a blue (metal-poor) halo and a red (metal-rich) halo, based on the geometric distinction between the BGC and RGC systems.
According to the velocity dispersion profiles, we conclude that the red halo of M85 traced by the RGC system is truncated at $R \sim 3-4 \arcmin$, corresponding to $\sim$ 18 kpc.

To investigate the extent of M85 halo, we compare the kinematics of the GCs with that of the dwarf galaxies in the Virgo Cluster, using radial velocities of dwarf galaxies presented in the EVCC \citep{kim14}.
There are only four Virgo dwarf galaxies in our survey region, and all of them have radial velocities higher than the M85 nucleus ({Figure \ref{fig:r_vr}}).
The mean radial velocity and radial velocity dispersion of these dwarf galaxies are \vrad = 1113 $\pm$ 158 \kms and $\sigma$ = 316 $\pm$ 112 \kmsend.

The kinematics of the M85 GCs is totally different from that of the Virgo dwarf galaxies in the survey region in terms of both the mean velocity and velocity dispersion.
This indicates that the gravitational potentials governing these two populations are different.
The mean radial velocity and velocity dispersion of the Virgo dwarf galaxies at the clustercentric distance same as M85 ($\sim 6\arcdeg$) are derived to be $\overline{v_{\rm r}}$ = 1107 \kms and $\sigma_{\rm r}$ = 327 \kmsend, respectively \citep{kim14}.
The dwarf galaxies in our survey region have kinematics similar to other dwarf galaxies in the Virgo despite the small sample.
Therefore, we conclude that the dwarf galaxies in our survey region follow the cluster potential, while the blue halo traced by the BGCs are controlled by the distinguishable galaxy potential.

	\subsection{Dynamical Mass in M85}

We estimated the dynamical mass of M85 based on the kinematics of the GC system.
The M85 GCs show clearly different kinematics according to their colors.
We only used the BGC system to derive the dynamical mass of M85 because it is a pressure-supported system that shows a negligible rotation feature. 

We estimate the pressure-supported mass of M85 using the tracer mass estimator (TME) from \citet{wea10}.
The TME is a robust method to estimate the enclosed mass based on the projected positions and line-of-sight velocities of tracers.
The enclosed mass based on the TME method is given by
\begin{equation}
	M_{\rm p} = \frac{C}{GN} \sum_{i}^{N} (v_{{\rm los},i}-v_{\rm sys})^2 ~R_{i}^{\alpha},
\end{equation}
where $N$ is the number of the tracers, $v_{{\rm los},i}$ is the rotation-corrected radial velocity of a given tracer, $v_{\rm sys}$ is the systemic velocity, $\alpha$ is the power-law slope of the underlying gravitational potential profile, and $R_{i}$ is the projected galactocentric distance of the tracers.
The constant $C$ is defined as
\begin{equation}
	C = \frac{\alpha+\gamma-2\beta}{I_{\alpha,\beta}}~r_{\rm out}^{1-\alpha},
\end{equation}
where $\gamma$ is the power-law slope of the volume number density profile of the tracers, $\beta$ is the anisotropy parameter ($\beta = 1 - \sigma_{\rm t}^2 / \sigma_{\rm r}^2$), $r_{\rm out}$ is the deprojected radius of the outermost tracer and
\begin{equation}
	I_{\alpha,\beta} = \frac{\pi^{1/2} \Gamma(\frac{\alpha}{2}+1)}{4 \Gamma(\frac{\alpha}{2}+\frac{5}{2})} ~[\alpha + 3 - \beta (\alpha + 2)].
\end{equation}

We adopted the $\gamma$ parameter of 3.28 from the number density profile of BGCs based on a wide-field photometric survey given by Paper I.
The $\alpha$ parameter is zero for the isothermal dark matter halo which shows a flat rotation curve and 0.55 for the NFW dark matter profile \citep{nfw96, wea10}.
The $\beta$ parameter is zero for isotropic orbits and one for purely radial orbits.
{We estimated the dynamical mass uncertainty from the bootstrapping method. We randomly selected 41 objects from the BGCs allowing replacement, and calculated the dynamical mass with their radial velocities based on the TME method. In this process, the radial velocities and the slope of the number density profile of BGCs are also randomly chosen within their uncertainties. We repeated this process 1000 times, and found the 16th and 84th percentiles of the measurements. We adopted the differences between the mass derived with the actual data and the 16th/84th percentiles as the uncertainty.}

We estimate the dynamical mass of M85 enclosed within $R$ = 124 kpc to be $M_{\rm TME} = (3.8 \pm 0.6) \times 10^{12} M_{\odot}$, assuming the isothermal dark matter halo and the isotropic orbits. 
Previously, \citet{san06} derived the mass of M85 enclosed within $R$ = 10 kpc from X-ray hot gas observation, which is $2 \times 10^{11} M_{\odot}$.
{In addition, \citet{bab18} estimated the total mass of M85 to be $4 \times 10^{11} M_{\odot}$ by extrapolating the total mass profiles out to 5 $R_{\rm e}$ (37 kpc). These measurements are 10-20 times smaller than the dynamical mass derived in this study. It is because M85 lacks hot gas, compared to other galaxies with similar luminosity \citep{san06}, and our radial coverage is much larger than the previous studies.}

We compare the dynamical mass of M85 to those of early-type galaxies with similar luminosity of M85.
\citet{ala16} presented dynamical masses of 23 early-type galaxies derived with their GC kinematics. Among them, we selected 7 galaxies with $K$-band magnitudes of --25.5 mag$< M_K <$ --24.5 mag, which are similar to that of M85 \citep[$M_K = -25.1$ mag;][]{jar03}.
These galaxies have pressure-supported masses ranging from $6.5 \times 10^{11} M_{\odot}$ to $3.26 \times 10^{12} M_{\odot}$ with radial coverage of $R < 9 - 21$ effective radii ($R_{\rm e}$). We derive the enclosed mass of M85 within $R = 14~R_{\rm e}$, comparable to the coverage for other galaxies, adopting the effective radius of M85, $R_{\rm e} = 102\farcs28$ \citep{kor09}.
M85 has a larger dynamical mass, compared to the other galaxies with similar luminosity, which indicates the existence of a large amount of dark matter in M85 within $R = 14~R_{\rm e}$.

\section{Summary}

We present a spectroscopic study of GCs in the merger remnant galaxy M85 using the MMT/Hectospec.
We identify 89 GCs in the radial range of $1\farcm5 < R < 31\arcmin$ based on radial velocity measurements.
We divided the confirmed GCs into three groups according to their colors: 41 BGCs with $0.6 < (g-i)_0 < 0.7$, 32 GGCs with $0.7 < (g-i)_0 < 0.8$, and 16 RGCs with $0.8 < (g-i)_0 < 1.0$.
The GC subpopulations show notable differences in their spatial distribution, kinematics, and mean metallicities. We could not find significant differences in their ages, showing that all GC subpopulations as old as 10 Gyr on average. 
The detailed properties of each GC subpopulation are summarized as follows.

\begin{itemize}

\item
The BGC system has the mean radial velocity of $\overline{v_{\rm r}} = 727^{+30}_{-29}$ \kmsend, slightly higher than the systemic velocity of M85 ($v_{\rm sys}$ = 696 \kmsend), and shows little rotation. 
The velocity dispersion of the BGCs is $\sigma_r = 168^{+14}_{-23}$ \kmsend.
The BGCs are the most metal-poor population among all GC subpopulations in M85, having the mean metallicity of [Z/H] = --1.49.

\item
Most of the GGCs have radial velocities much higher than the systemic velocity of M85 with the mean radial velocity of $\overline{v_{\rm r}} = 812^{+30}_{-27}$ \kmsend.
They constitutes a stream out to $R = 31\arcmin$ along the major axis of the outer isophotes of M85.
The GGCs have the mean metallicity of [Z/H] = --0.91, which are between those of the BGCs and RGCs.

\item
The mean radial velocity of the RGCs is $\overline{v_{\rm r}} = 704 \pm 37$ \kmsend, consistent with the systemic velocity.
The RGC system shows a disk-like strong rotation with the rotation parameter of $\Omega R/\sigma_{\rm r, cor} = 2.15$.
The rotation-corrected velocity dispersion of the RGCs within $R = 6\arcmin$ is $\sigma_{\rm r, cor}$ = 66 \kmsend, much smaller than that of the BGCs and the central stars.
The mean metallicity of the RGCs is [Z/H] = --0.45, which is highest among those of all GC subpopulations in M85.

\end{itemize}

These differences in the kinematics of the GC subpopulations imply that they have different formation and evolution histories.
The BGCs in M85 have kinematic properties similar to those in other massive early-type galaxies, which are expected to be accreted from the disrupted dwarf galaxies. The metal-poor population of the BGCs also supports this scenario.
On the other hand, the GGCs and RGCs in M85 have peculiar kinematics that cannot be explained by the typical GC formation scenarios.
The GGCs may be a population accreting to or escaping from the M85 plane, and the RGCs may be a remnant produced by recent off-center major merging events.
Comparing their spatial distribution and kinematics with those of planetary nebulae in the outer stellar light would be helpful to understand the origin of these GCs. 

In addition, we investigate the extent and dynamical mass of the M85 halo using the GC kinematics.
The mean radial velocity and velocity dispersion of the GCs in M85 are different from those of Virgo dwarf galaxies around M85.
Especially, the low velocity dispersion of the RGC system indicates a truncation of the red halo of M85.
The BGCs are distributed out to $R = 23\farcm9$, having the velocity dispersion lower than Virgo dwarf galaxies.
Therefore, we conclude that M85 has a distinguishable galaxy potential at least out to $R = 23\farcm9$ corresponding to 124 kpc.
We derive the dynamical mass of M85 using the kinematics of the pressure-supported BGC system to be 3.8 $\times$ 10$^{12} M_{\odot}$, assuming the isothermal dark matter halo and the isotropic orbit.

\acknowledgments
We thank Brian S. Cho for his helpful comments in improving the English of this manuscript.
This project was supported by the National Research Foundation grant funded by the Korean Government (NRF-2019R1A2C2084019) and the K-GMT Science Program (PID: 2016A-UAO-G4; 2016A-MMT-3) funded through Korean GMT Project operated by Korea Astronomy and Space Science Institute (KASI).
H.S.P. was supported in part by the National Research Foundation of Korea (NRF) grant funded by the Korea government (MSIT, Ministry of Science and ICT; No. NRF-2019R1F1A1058228).
Observations reported here were obtained at the MMT Observatory, a joint facility of the University of Arizona and the Smithsonian Institution.

\end{document}